\documentclass[12pt, reqno]{amsart}

\usepackage{amssymb, amsmath}
\usepackage[all]{xy}

\newtheorem{thm1}{Theorem}[section]
\newtheorem{lem1}[thm1]{Lemma}

\newtheorem{def1}[thm1]{Definition}

\newtheorem{prop1}[thm1]{Proposition}

\begin{document}

\title[BOSONISATION AND PARASTATISTICS]
{BOSONISATION AND PARASTATISTICS: AN EXAMPLE AND AN ALTERNATIVE
APPROACH}
\author[K. Kanakoglou, C. Daskaloyannis]{K. Kanakoglou, C. Daskaloyannis}
\address {K. Kanakoglou: Department of Physics, Aristotle University of Thessaloniki, Thessaloniki 54124, GREECE
 } \email{kanakoglou@hotmail.com}
\address {C. Daskaloyannis: Department of Mathematics, Aristotle University of Thessaloniki, Thessaloniki 54124, GREECE
 } \email{daskalo@math.auth.gr}

\begin{abstract}
Definitions of the parastatistics algebras and known results on
their Lie (super)algebraic structure are reviewed. The notion of
super-Hopf algebra is discussed. The bosonisation technique for
switching a Hopf algebra in a braided category ${}_{H}\mathcal{M}$
($H$: a quasitriangular Hopf algebra) into an ordinary Hopf
algebra is presented and it is applied in the case of the
parabosonic algebra. A bosonisation-like construction is also
introduced for the same algebra and the differences are discussed.
\end{abstract}

\maketitle

\section{Introduction and Definitions}

Throughout this paper we are going to use the following notation
conventions: \\
If $x$ and $y$ are any monomials of the tensor algebra of some
k-vector space, we are going to call commutator the following
expression:
$$
[x,y] = x \otimes y - y \otimes x \equiv xy - yx
$$
and anticommutator the following expression:
$$
\{x,y\} = x \otimes y + y \otimes x \equiv xy + yx
$$
By the field $k$ we shall always mean $\mathbb{C}$, and all tensor
products will be considered over $k$ unless stated so. Finally we
freely use Sweedler's notation for the comultiplication throughout
the paper.

Parafermionic and parabosonic algebras first appeared in the
physics literature by means of generators and relations, in the
pionnering works of Green \cite{Green} and Greenberg and Messiah
\cite{GreeMe}. Their purpose was to introduce generalizations of
the usual bosonic and fermionic algebras of quantum mechanics,
capable of leading to generalized versions of the Bose-Einstein
and Fermi-Dirac statistics (see: \cite{OhKa}). We start with the
definitions of these algebras:

 Let us consider the k-vector space $V_{B}$ freely generated by
  the elements: $b_{i}^{+}, b_{j}^{-}$, $i,j=1,...,n$. Let
$T(V_{B})$ denote the tensor algebra of $V_{B}$ (i.e.: the free
algebra generated by the elements of the basis). In $T(V_{B})$ we
consider the two-sided ideal $I_{B}$ generated by the following
elements:
\begin{equation} \label{eq:pbdef}
 \big[ \{ b_{i}^{\xi},  b_{j}^{\eta}\}, b_{k}^{\epsilon}  \big] -
 (\epsilon - \eta)\delta_{jk}b_{i}^{\xi} - (\epsilon - \xi)\delta_{ik}b_{j}^{\eta}
\end{equation}
for all values of $\xi, \eta, \epsilon = \pm 1$ and $i,j,k = 1,
\ldots, n$. \\
We now have the following:

\begin{def1}
The parabosonic algebra in $2n$ generators $P_{B}^{(n)}$ ($n$
parabosons) is the quotient algebra of the tensor algebra of
$V_{B}$ with the ideal $I_{B}$:
$$
P_{B}^{(n)} = T(V_{B}) / I_{B}
$$
\end{def1}

In a similar way we may describe the parafermionic algebra in $2n$
generators ($n$ parafermions): Let us consider th k-vector space
$V_{F}$ freely generated by the elements: $f_{i}^{+}, f_{j}^{-}$,
$i,j=1,...,n$. Let $T(V_{F})$ denote the tensor algebra of $V_{F}$
(i.e.: the free algebra generated by the elements of the basis).
In $T(V_{F})$ we consider the two-sided ideal $I_{F}$ generated by
the following elements:

\begin{equation} \label{eq:pfdef}
\big[ [ f_{i}^{\xi},  f_{j}^{\eta} ], f_{k}^{\epsilon}  \big] -
\frac{1}{2}(\epsilon - \eta)^{2}\delta_{jk}f_{i}^{\xi} +
\frac{1}{2}(\epsilon - \xi)^{2}\delta_{ik}f_{j}^{\eta}
\end{equation}
for all values of $\xi, \eta, \epsilon = \pm 1$ and $i,j,k = 1,
\ldots, n$. \\
We get the following definition:

\begin{def1}
The parafermionic algebra in $2n$ generators $P_{F}^{(n)}$ ($n$
parafermions) is the quotient algebra of the tensor algebra of
$V_{F}$ with the ideal $I_{F}$:
$$
P_{F}^{(n)} = T(V_{F}) / I_{F}
$$
\end{def1}

\section{(super-)Lie and (super-)Hopf algebraic structure of $P_{B}^{(n)}$ and $P_{F}^{(n)}$}

Due to it's simpler nature, parafermionic algebras were the first
to be identified as the universal enveloping algebras (UEA) of
simple Lie algebras. This was done  almost at the same time by
S.Kamefuchi, Y.Takahashi in \cite{Kata} and by C. Ryan, E.C.G.
Sudarshan in \cite{RySu}. In fact the following stem from the
above mentioned references (see also \cite{Pal1}):

\begin{lem1}
In the k-vector space $P_{F}^{(n)}$ we consider the  k-subspace
generated by the set of elements:
$$
 \Big\{ [f_{i}^{\xi}, f_{j}^{\eta}], f_{k}^{\epsilon} \ \ |
\xi, \eta, \epsilon = \pm, \ i,j,k = 1,\ldots,n \Big\}
$$
The above subspace endowed with a bilinear multiplication
$\langle..,..\rangle$ whose values are determined by the values of
the commutator in $P_{F}^{(n)}$, i.e:
$$
\langle f_{i}^{\xi}, f_{j}^{\eta} \rangle = [f_{i}^{\xi},
f_{j}^{\eta}]
$$
and:
$$
\big\langle [ f_{i}^{\xi}, f_{j}^{\eta} ], f_{k}^{\epsilon}
\big\rangle = \big[ [ f_{i}^{\xi}, f_{j}^{\eta} ],
f_{k}^{\epsilon} \big] = \frac{1}{2}(\epsilon -
\eta)^{2}\delta_{jk}f_{i}^{\xi} - \frac{1}{2}(\epsilon -
\xi)^{2}\delta_{ik}f_{j}^{\eta}
$$
is a simple complex Lie algebra isomorphic to $B_{n} = so(2n+1)$.
The basis in the Cartan subalgebra of $B_{n}$ can be chosen in
such a way that the elements $f^{+}$ (respectively: $f^{-}$) are
negative (respectively: positive) root vectors.
\end{lem1}

Based on the above observations, the following is finally proved:

\begin{prop1}
The parafermionic algebra in $2n$ generators is isomorphic to the
universal enveloping algebra of the simple complex Lie algebra
$B_{n} = so(2n+1)$ (according to the well known classification of
the simple complex Lie algebras), i.e:
$$
P_{F}^{(n)} \cong U(B_{n})
$$
\end{prop1}

An immediate consequence of the above identification is that
parafermionic algebras are ordinary Hopf algebras, with the
generators $f_{i}^{\pm}$, $i=1,...,n$ being primitive elements.
The Hopf algebraic structure of $P_{F}^{(n)}$ is completely
determined by the well known Hopf algebraic structure of the Lie
algebras, due to the above isomorphism. For convenience we quote
the relations explicitly:
\begin{equation} \label{eq:HopfPF}
\begin{array}{ccccc}
  \Delta(f_{i}^{\pm}) = f_{i}^{\pm} \otimes 1 + 1 \otimes f_{i}^{\pm} & & \varepsilon(f_{i}^{\pm}) = 0
  & &  S(f_{i}^{\pm}) = -f_{i}^{\pm}  \\
\end{array}
\end{equation}

The algebraic structure of parabosons seemed to be somewhat more
complicated. The presence of anticommutators among the trilinear
 relations defining $P_{B}^{(n)}$ ``breaks'' the usual (Lie)
antisymmetry and makes impossible the identification of the
parabosons with the UEA of any Lie algebra. It was in the early
'80 's that was conjectured \cite{OhKa}, that due to the mixing of
commutators and anticommutators in $P_{B}^{(n)}$ the proper
mathematical ``playground" should be some kind of Lie superalgebra
(or: $\mathbb{Z}_{2}$-graded Lie algebra). Starting in the early
'80 's, and using the recent (by that time) results in the
classification of the finite dimensional simple complex Lie
superalgebras which was obtained by Kac (see: \cite{Kac1, Kac2}),
T.D.Palev managed to identify the parabosonic algebra with the UEA
of a certain simple complex Lie superalgebra. In \cite{Pal3},
\cite{Pal5} (see also \cite{Pal2}), T.D.Palev shows the following:

\begin{lem1}
In the k-vector space $P_{B}^{(n)}$ we consider the  k-subspace
generated by the set of elements:
$$
\Big\{ \{b_{i}^{\xi}, b_{j}^{\eta}\}, b_{k}^{\epsilon} \ \ | \xi,
\eta, \epsilon = \pm, \ i,j,k = 1,\ldots,n \Big\}
$$
This vector space is turned into a superspace (
$\mathbb{Z}_{2}$-graded vector space ) by the requirement that
$b_{i}^{\xi}$ span the odd subspace and $\{b_{i}^{\xi},
b_{j}^{\eta}\}$ span the even subspace. \\
 The above vector space endowed with a bilinear multiplication $\langle..,..\rangle$
 whose values are determined by the values of
the anticommutator and the commutator in $P_{B}^{(n)}$, i.e.:
$$
\langle b_{i}^{\xi}, b_{j}^{\eta} \rangle = \{b_{i}^{\xi},
b_{j}^{\eta}\}
$$
and:
$$
\big\langle \{ b_{i}^{\xi},  b_{j}^{\eta}\}, b_{k}^{\epsilon}
\big\rangle = \big[ \{ b_{i}^{\xi},  b_{j}^{\eta}\},
b_{k}^{\epsilon} \big] = (\epsilon - \eta)\delta_{jk}b_{i}^{\xi} +
(\epsilon - \xi)\delta_{ik}b_{j}^{\eta}
$$
respectively, according to the above mentioned gradation, is a
simple, complex super-Lie algebra (or: $\mathbb{Z}_{2}$-graded Lie
algebra) isomorphic to $B(0,n) = osp(1,2n)$. The basis in the
Cartan subalgebra of $B(0,n)$ can be chosen in such a way that the
elements $b^{+}$ (respectively: $b^{-}$) are negative
(respectively: positive) root vectors.
\end{lem1}

Note that, according to the above lemma, the even part of $B(0,n)$
is spanned by the elements $\big\{ \{b_{i}^{\xi}, b_{j}^{\eta}\} \
\ | \xi, \eta = \pm, \ i,j = 1,\ldots,n \big\}$ and is a
subalgebra of $B(0,n)$ isomorphic to the Lie algebra $sp(2n)$.
It's Lie multiplication can be readily deduced from the above
given commutators and reads:
$$
\begin{array}{c}
 \big\langle \{ b_{i}^{\xi},  b_{j}^{\eta}\}, \{ b_{k}^{\epsilon},
b_{l}^{\phi} \} \big\rangle = \big[ \{ b_{i}^{\xi},
b_{j}^{\eta}\}, \{ b_{k}^{\epsilon}, b_{l}^{\phi} \} \big] =
  \\
      \\
  (\epsilon - \eta)\delta_{jk} \{ b_{i}^{\xi}, b_{l}^{\phi} \} +
(\epsilon - \xi)\delta_{ik}\{ b_{j}^{\eta}, b_{l}^{\phi} \} +
(\phi - \eta)\delta_{jl}\{ b_{i}^{\xi}, b_{k}^{\epsilon} \} +
(\phi - \xi)\delta_{il}\{ b_{j}^{\eta}, b_{k}^{\epsilon} \} \\
\end{array}
$$
On the other hand the elements $\big\{ b_{k}^{\epsilon} \ \ |
\epsilon = \pm, \ k = 1,\ldots,n \big\}$ constitute a basis of the
odd part of $B(0,n)$.

 Note also, that $B(0,n)$  in Kac's
notation, is the classical simple complex orthosymplectic Lie
superalgebra denoted $osp(1,2n)$ in the notation traditionally
used by physicists until then.

Based on the above observations, Palev finally proves (in the
above mentioned references):

\begin{prop1} \label{parab}
The parabosonic algebra in $2n$ generators is isomorphic to the
universal enveloping algebra of the classical simple complex Lie
superalgebra $B(0,n)$ (according to the classification of the
simple complex Lie superalgebras given by Kac), i.e:
$$
P_{B}^{(n)} \cong U(B(0,n))
$$
\end{prop1}

 The universal enveloping algebra $U(L)$ of a Lie superalgebra
$L$ is not a Hopf algebra, at least in the ordinary sense. $U(L)$
is a $\mathbb{Z}_{2}$-graded associative algebra (or:
superalgebra) and it is a super-Hopf algebra in a sense that we
briefly describe: First we consider the braided tensor product
algebra $U(L) \underline{\otimes} U(L)$, which means the vector
space $U(L) \otimes U(L)$ equipped with the associative
multiplication:
$$
(a \otimes b) \cdot (c \otimes d) = (-1)^{|b||c|}ac \otimes bd
$$
for $b,c$ homogeneous elements of $U(L)$, and $\ |.| \ $ denotes
the degree of an homogeneous element (i.e.: $|b|=0$ if $b$ is an
even element and $|b|=1$ if $b$ is an odd element). Note that
$U(L) \underline{\otimes} U(L)$ is also a superalgebra or:
$\mathbb{Z}_{2}$-graded associative algebra. Then $U(L)$ is
equipped with a coproduct
$$
\underline{\Delta} : U(L) \rightarrow U(L) \underline{\otimes}
U(L)
$$
which is an superalgebra homomorphism from $U(L)$ to the braided
tensor product algebra  $U(L) \underline{\otimes} U(L)$ :
$$
\underline{\Delta}(ab) = \sum (-1)^{|a_{2}||b_{1}|}a_{1}b_{1}
\otimes a_{2}b_{2} = \underline{\Delta}(a) \cdot
\underline{\Delta}(b)
$$
for any $a,b$ in $U(L)$, with $\Delta(a) = \sum a_{1} \otimes
a_{2}$, $\Delta(b) = \sum b_{1} \otimes b_{2}$, and $a_{2}$,
$b_{1}$ homogeneous.  $\underline{\Delta}$ is uniquely determined
by it's value on the generators of $U(L)$ (i.e.: the basis
elements of $L$):
$$
\underline{\Delta}(x) = 1 \otimes x + x \otimes 1
$$
 Similarly, $U(L)$ is equipped with an antipode $\underline{S} : U(L)
\rightarrow U(L)$ which is not an algebra anti-homomorphism (as in
ordinary Hopf algebras) but a braided algebra anti-homomorphism
(or: ``twisted" anti-homomorphism) in the following sense:
$$
\underline{S}(ab) = (-1)^{|a||b|}\underline{S}(b)\underline{S}(a)
$$
for any homogeneous $a,b \in U(L)$. \\

All the above description is equivalent to saying that $U(L)$ is a
Hopf algebra in the braided category of $\mathbb{CZ}_{2}$-modules
${}_{\mathbb{CZ}_{2}}\mathcal{M}$ or: a braided group where the
braiding is induced by the non-trivial quasitriangular structure
of the $\mathbb{CZ}_{2}$ Hopf algebra i.e. by the non-trivial
$R$-matrix:
\begin{equation} \label{eq:nontrivRmatrcz2}
R_{g} = \frac{1}{2}(1 \otimes 1 + 1 \otimes g + g \otimes 1 - g
\otimes g)
\end{equation}
where $1, g$ are the elements of the $\mathbb{Z}_{2}$ group which
is now written multiplicatively. \\
We recall here (see \cite{Mon}) that if $(H,R_{H})$ is a
quasitriangular Hopf algebra, then the category of modules
${}_{H}\mathcal{M}$ is a braided monoidal category, where the
braiding is given by a natural family of isomorphisms $\Psi_{V,W}:
V \otimes W \cong W \otimes V$, given explicitly by:
\begin{equation} \label{eq:braid}
\Psi_{V,W}(v \otimes w) = \sum (R_{H}^{(2)} \vartriangleright w)
\otimes (R_{H}^{(1)} \vartriangleright v)
\end{equation}
for any $V,W \in obj({}_{H}\mathcal{M})$.   \\
Combining eq. \eqref{eq:nontrivRmatrcz2} and \eqref{eq:braid} we
immediately get the braiding in the
${}_{\mathbb{CZ}_{2}}\mathcal{M}$ category:
\begin{equation} \label{symmbraid}
\Psi_{V,W}(v \otimes w) = (-1)^{|v||w|} w \otimes v
\end{equation}
This is obviously a symmetric braiding, so we actually have a
symmetric monoidal category ${}_{\mathbb{CZ}_{2}}\mathcal{M}$,
rather than a truly braided one. \\

In view of the above description, an immediate consequence of
proposition \ref{parab}, is that the parabosonic algebras
$P_{B}^{(n)}$ are super-Hopf algebras, with the generators
$b_{i}^{\pm}$, $i=1,...,n$ being primitive elements. It's
super-Hopf algebraic structure is completely determined by the
super-Hopf algebraic structure of Lie superalgebras, due to the
above mentioned isomorphism. Namely the following relations
determine completely the super-Hopf algebraic structure of
$P_{B}^{(n)}$:
\begin{equation} \label{eq:HopfPB}
\begin{array}{ccccc}
  \underline{\Delta}(b_{i}^{\pm}) = 1 \otimes b_{i}^{\pm} + b_{i}^{\pm} \otimes 1 &
  & \underline{\varepsilon}(b_{i}^{\pm}) = 0  & & \underline{S}(b_{i}^{\pm}) = - b_{i}^{\pm} \\
\end{array}
\end{equation}

\section{Bosonisation as a technique of reducing supersymmetry}

A general scheme for ``transforming" a Hopf algebra $B$ in the
braided category ${}_{H}\mathcal{M}$ ($H$: some quasitriangular
Hopf algebra) into an ordinary one, namely the smash product Hopf
algebra: $B \star H$, such that the two algebras have equivalent
module categories, has been developed during '90 's. The original
reference is \cite{Maj1} (see also \cite{Maj2, Maj3}). The
technique is called bosonisation, the term coming from physics.
This technique uses ideas developed in \cite{Ra}, \cite{Mo}. It is
also presented and applied in \cite{Fi}, \cite{FiMon},
\cite{Andru}. We review the main points of the above method:

 In general, $B$ being a Hopf algebra in a category, means
that it's structure maps are morphisms in the category. In
particular, if $H$ is some quasitriangular Hopf algebra, B being a
Hopf algebra in the braided category ${}_{H}\mathcal{M}$ , means
that $B$ is an algebra in ${}_{H}\mathcal{M}$ (or: $H$-module
algebra) and a coalgebra in ${}_{H}\mathcal{M}$ (or: $H$-module
coalgebra) and at the same time $\Delta_{B}$ and $\varepsilon_{B}$
are algebra morphisms in the category ${}_{H}\mathcal{M}$. (For more details
on the above definitions one may consult for example \cite{Mon}). \\
Since $B$ is an $H$-module algebra we can form the cross product
algebra $B \rtimes H$ (also called: smash product algebra) which
as a k-vector space is $B \otimes H$ (i.e. we write: $b \rtimes h
\equiv b \otimes h$ for every $b \in B$, $h \in H$), with
multiplication given by:
\begin{equation} \label{eq:crosspralg}
(b \otimes h)(c \otimes g) = \sum b(h_{1} \vartriangleright c)
\otimes h_{2}g
\end{equation}
$\forall$ $b,c \in B$ and $h,g \in H$, and the usual tensor
product unit. \\
On the other hand $B$ is a (left) $H$-module coalgebra with $H$:
quasitriangular through the $R$-matrix: $R_{H} = \sum R_{H}^{(1)}
\otimes R_{H}^{(2)}$.
 Quasitriangularity
``switches" the (left) action of $H$ on $B$ into a (left) coaction
$\rho: B \rightarrow H \otimes B$ through:
\begin{equation} \label{eq:act-coact}
\rho(b) = \sum R_{H}^{(2)} \otimes (R_{H}^{(1)} \vartriangleright
b)
\end{equation}
and $B$ endowed with this coaction becomes (see \cite{Maj2, Maj3})
a (left) $H$-comodule coalgebra or equivalently a coalgebra in
${}^{H}\mathcal{M}$ (meaning that $\Delta_{B}$ and
$\varepsilon_{B}$ are (left) $H$-comodule morphisms, see \cite{Mon}). \\
We recall here (see: \cite{Maj2, Maj3}) that when $H$ is a Hopf
algebra and $B$ is a (left) $H$-comodule coalgebra with the (left)
$H$-coaction given by: $\rho(b) = \sum b^{(1)} \otimes b^{(0)}$ ,
one may form the cross coproduct coalgebra $B \rtimes H$, which as
a k-vector space is $B \otimes H$ (i.e. we write: $b \rtimes h
\equiv b \otimes h$ for every $b \in B$, $h \in H$), with
comultiplication given by:
\begin{equation} \label{eq:crosscoprcoalg}
\Delta(b \otimes h) = \sum b_{1} \otimes b_{2}^{(1)}h_{1} \otimes
b_{2}^{(0)} \otimes h_{2}
\end{equation}
and counit: $\varepsilon(b \otimes h) = \varepsilon_{B}(b)
\varepsilon_{H}(h)$. (In the above we use in the elements of $B$
upper indices included in parenthesis to denote the components of
the coaction according to the Sweedler notation, with the
convention that $b^{(i)} \in H$
for $i \neq 0$). \\
Now we proceed by applying the above described construction of the
cross coproduct coalgebra $B \rtimes H$ , with the special form of
the (left) coaction given by eq. \eqref{eq:act-coact}. Replacing
thus eq. \eqref{eq:act-coact} into eq. \eqref{eq:crosscoprcoalg}
we get for the special case of the quasitriangular Hopf algebra H
the cross coproduct comultiplication:
\begin{equation} \label{eq:crosscoprcoalgR}
\Delta(b \otimes h) = \sum b_{1} \otimes R_{H}^{(2)}h_{1} \otimes
(R_{H}^{(1)} \vartriangleright b_{2}) \otimes h_{2}
\end{equation}
Finally we can show that the cross product algebra (with
multiplication given by \eqref{eq:crosspralg}) and the cross
coproduct coalgebra (with comultiplication given by
\eqref{eq:crosscoprcoalgR}) fit together and form a bialgebra
(see: \cite{Maj2, Maj3, Mo, Ra}). This bialgebra, furnished with
an antipode:
\begin{equation} \label{antipodecrosspr}
S(b \otimes h) = (S_{H}(h_{2}))u(R^{(1)} \vartriangleright
S_{B}(b)) \otimes S(R^{(2)}h_{1})
\end{equation}
where $u = \sum S_{H}(R^{(2)})R^{(1)}$, and $S_{B}$ the (braided)
antipode of $B$, becomes (see \cite{Maj2}) an ordinary Hopf
algebra. This is the smash product Hopf algebra denoted $B \star
H$. In \cite{Maj1} it is further proved that the category of the
braided modules of $B$ ($B$-modules in ${}_{H}\mathcal{M}$) is
equivalent to the category of the (ordinary) modules of $B \star
H$.

   \subsection{An example of Bosonisation}
 In the special case that $B$ is some super-Hopf
algebra, then: $H = \mathbb{CZ}_{2}$, equipped with it's
non-trivial quasitriangular structure, formerly mentioned. In this
case, the technique simplifies and the ordinary Hopf algebra
produced is the smash product Hopf algebra $B \star
\mathbb{CZ}_{2}$. The grading in $B$ is induced by the
$\mathbb{CZ}_{2}$-action on $B$:
\begin{equation} \label{eq:cz2action}
g \vartriangleright b = (-1)^{|b|}b
\end{equation}
for $b$ homogeneous in $B$. Utilizing the non-trivial $R$-matrix
$R_{g}$ and using eq. \eqref{eq:nontrivRmatrcz2} and eq.
\eqref{eq:act-coact} we can readily deduce the form of the induced
$\mathbb{CZ}_{2}$-coaction on $B$:
\begin{equation} \label{eq:cz2coaction}
\rho(b) = \left\{ \begin{array}{ccc}
    1 \otimes b & , & b: \textrm{even} \\
    g \otimes b & , & b: \textrm{odd} \\
\end{array} \right.
\end{equation}
The above mentioned action and coaction enable us to form the
cross product algebra and the cross coproduct coalgebra according
to the preceding discussion which finally form the smash product
Hopf algebra $B \star \mathbb{CZ}_{2}$.  The grading of $B$, is
``absorbed" in $B \star \mathbb{CZ}_{2}$, and becomes an inner
automorphism:
$$
gbg = (-1)^{|b|}b
$$
where we have identified: $b \star 1 \equiv b$ and $1 \star g
\equiv g$ in $B \star \mathbb{CZ}_{2}$ and $b$ homogeneous element
in $B$. This inner automorphism is exactly the adjoint action of
$g$ on $B \star \mathbb{CZ}_{2}$ (as an ordinary Hopf algebra).
The following proposition is proved -as an example of the
bosonisation technique- in \cite{Maj2}:
\begin{prop1} \label{bosonisat}
Corresponding to every super-Hopf algebra $B$ there is an ordinary
Hopf algebra $B \star \mathbb{CZ}_{2}$, its bosonisation,
consisting of $B$ extended by adjoining an element $g$ with
relations, coproduct, counit and antipode:
\begin{equation} \label{eq:HopfPBg}
\begin{array}{cccc}
  g^{2} = 1 & gb = (-1)^{|b|}bg & \Delta(g) = g \otimes g & \Delta(b) = \sum b_{1}g^{|b_{2}|} \otimes b_{2} \\
                                     \\
  S(g) = g & S(b) = g^{-|b|}\underline{S}(b) & \varepsilon(g) = 1 & \varepsilon(b) = \underline{\varepsilon}(b) \\
\end{array}
\end{equation}
where $\underline{S}$ and $\underline{\varepsilon}$ denote the
original maps of the super-Hopf algebra $B$. \\
Moreover, the representations of the bosonised Hopf algebra $B
\rtimes \mathbb{CZ}_{2}$ are precisely the super-representations
of the original superalgebra $B$.
\end{prop1}
The application of the above proposition in the case of the
parabosonic algebra $P_{B}^{(n)} \cong U(B(0,n))$ is
straightforward: we immediately get it's bosonised form
$P_{B(g)}^{(n)}$ which by definition is:
$$
P_{B(g)}^{(n)} \equiv P_{B}^{(n)} \star \mathbb{CZ}_{2} \cong
U(B(0,n)) \star \mathbb{CZ}_{2}
$$
Utilizing equations \eqref{eq:HopfPB} which describe the
super-Hopf algebraic structure of the parabosonic algebra
$P_{B}^{(n)}$, and replacing them into equations
\eqref{eq:HopfPBg} which describe the ordinary Hopf algebra
structure of the bosonised superalgebra, we immediately get the
explicit form of the (ordinary) Hopf algebra structure of
$P_{B(g)}^{(n)} \equiv P_{B}^{(n)} \star \mathbb{CZ}_{2}$ which
reads:
\begin{equation} \label{eq:HopfPBgexpl}
\begin{array}{cc}
  \Delta(b_{i}^{\pm}) = b_{i}^{\pm} \otimes 1 + g \otimes b_{i}^{\pm} & \Delta(g) = g \otimes g \\
       \\
  \varepsilon(b_{i}^{\pm}) = 0 & \varepsilon(g) = 1 \\
                     \\
  S(b_{i}^{\pm}) = b_{i}^{\pm}g = -gb_{i}^{\pm} & S(g) = g \\
                      \\
  g^{2} = 1 & \{g,b_{i}^{\pm}\} = 0  \\
\end{array}
\end{equation}
where we have again identified $b_{i}^{\pm} \star 1 \equiv
b_{i}^{\pm}$ and $1 \star g \equiv g$ in $P_{B}^{(n)} \star
\mathbb{CZ}_{2}$.

\subsection{An alternarive approach}

Let us describe now a slightly different construction (see:
\cite{DaKa}), which achieves the same object: the determination of
an ordinary Hopf structure for the parabosonic algebra
$P_{B}^{(n)}$. \\
 Defining:
$$
N_{lm} = \frac{1}{2}\{b_{l}^{+},b_{m}^{-}\}
$$
we notice that these are the generators of the Lie algebra $u(n)$:
$$
[N_{kl},N_{mn}] = \delta_{lm}N_{kn} - \delta_{kn}N_{ml}
$$
We introduce now the elements:
$$
\mathcal{N} = \sum_{i=1}^{n}N_{ii} =
\frac{1}{2}\sum_{i=1}^{n}\{b_{i}^{+},b_{i}^{-}\}
$$
which are exactly the linear Casimirs of $u(n)$. \\
We can easily find that they satisfy:
$$
[\mathcal{N}, b_{i}^{\pm}] = \pm b_{i}^{\pm}
$$
Based on the above we inductively prove:
\begin{equation} \label{eq:Casimcom}
[\mathcal{N}^{m}, b_{i}^{+}] = b_{i}^{+}((\mathcal{N} + 1)^{m} -
\mathcal{N}^{m})
\end{equation}
We now introduce the following elements:
$$
K^{+} = \exp(i \pi \mathcal{N}) \equiv \sum_{m=0}^{\infty}
\frac{(i \pi \mathcal{N})^{m}}{m!}
$$
and:
$$
K^{-} = \exp(-i \pi \mathcal{N}) \equiv \sum_{m=0}^{\infty}
\frac{(-i \pi \mathcal{N})^{m}}{m!}
$$
Utilizing the above power series expressions and equation
\eqref{eq:Casimcom} we get
\begin{equation} \label{eq:Kb}
\begin{array}{lr}
 \{K^{+},b_{i}^{\pm}\} = 0 &  \{K^{-},b_{i}^{\pm}\} = 0 \\
\end{array}
\end{equation}
A direct application of the Baker-Campbell-Hausdorff formula leads
also to:
\begin{equation} \label{eq:KK}
K^{+}K^{-} = K^{-}K^{+} = 1
\end{equation}
We finally have the following proposition:
\begin{prop1} \label{altern}
Corresponding to the super-Hopf algebra $P_{B}^{(n)}$ there is an
ordinary Hopf algebra $P_{B(K^{\pm})}^{(n)}$, consisting of
$P_{B}^{(n)}$ extended by adjoining two elements $K^{+}$, $K^{-}$
with relations, coproduct, counit and antipode:
\begin{equation} \label{eq:HopfPBK}
\begin{array}{cc}
  \Delta(b_{i}^{\pm}) = b_{i}^{\pm} \otimes 1 + K^{\pm} \otimes b_{i}^{\pm} & \Delta(K^{\pm}) = K^{\pm} \otimes K^{\pm} \\
       \\
  \varepsilon(b_{i}^{\pm}) = 0 & \varepsilon(K^{\pm}) = 1 \\
                     \\
  S(b_{i}^{\pm}) = b_{i}^{\pm}K^{\mp} & S(K^{\pm}) = K^{\mp} \\
                      \\
  K^{+}K^{-} = K^{-}K^{+} = 1 & \{K^{+},b_{i}^{\pm}\} = 0 = \{K^{-},b_{i}^{\pm}\} \\
\end{array}
\end{equation}
\end{prop1}

\begin{proof}
Consider the k-vector space $k\langle b_{i}^{+}, b_{j}^{-},
K^{\pm} \rangle$ freely generated by the elements $b_{i}^{+},
b_{j}^{-}, K^{+}, K^{-}$.  Denote $T(b_{i}^{+}, b_{j}^{-},
K^{\pm})$ its tensor algebra. In the tensor algebra we denote
$I_{BK}$ the ideal generated by al the elements of the forms
\eqref{eq:pbdef}, \eqref{eq:Kb}, \eqref{eq:KK}. We define:
$$
P_{B(K^{\pm})}^{(n)} = T(b_{i}^{+}, b_{j}^{-}, K^{\pm}) / I_{BK}
$$
Consider the k-linear map
$$
\Delta : k\langle b_{i}^{+}, b_{j}^{-}, K^{\pm} \rangle
\rightarrow P_{B(K^{\pm})}^{(n)} \otimes P_{B(K^{\pm})}^{(n)}
$$
determined by it's values on the basis elements, specified in
equation \eqref{eq:HopfPBK}. By the universality property of the
tensor algebra this map extends to an algebra homomorphism:
$$
\Delta: T(b_{i}^{+}, b_{j}^{-}, K^{\pm}) \rightarrow
P_{B(K^{\pm})}^{(n)} \otimes P_{B(K^{\pm})}^{(n)}
$$
Now we can trivially verify that
\begin{equation} \label{eq:DKb}
 \Delta(\{K^{\pm},b_{i}^{\pm}\})
= \Delta(K^{+}K^{-} -1) = \Delta(K^{-}K^{+}-1) = 0
\end{equation}
Considering the usual tensor product
algebra $P_{B(K^{\pm})}^{(n)} \otimes P_{B(K^{\pm})}^{(n)}$ with
multiplication $(a \otimes b)(c \otimes d) = ac \otimes bd$ for
any $a,b,c,d \in P_{B(K^{\pm})}^{(n)}$ we also compute:
\begin{equation} \label{eq:Db}
\Delta(\big[ \{ b_{i}^{\xi},  b_{j}^{\eta}\}, b_{k}^{\epsilon}
\big] -
 (\epsilon - \eta)\delta_{jk}b_{i}^{\xi} - (\epsilon -
 \xi)\delta_{ik}b_{j}^{\eta}) = 0
\end{equation}
Relations \eqref{eq:DKb}, and \eqref{eq:Db}, mean that $I_{BK}
\subseteq ker \Delta$ which in turn implies that $\Delta$ is
uniquely extended as an algebra homomorphism from $
P_{B(K^{\pm})}^{(n)}$ to  the usual tensor product algebra
$P_{B(K^{\pm})}^{(n)} \otimes P_{B(K^{\pm})}^{(n)}$ according to
the diagram:
\begin{displaymath}
\xymatrix{T(b_{i}^{+}, b_{j}^{-}, K^{\pm}) \ar[rr]^{\Delta}
\ar[dr]_{\pi} & & P_{B(K^{\pm})}^{(n)} \otimes
P_{B(K^{\pm})}^{(n)} \\
 & P_{B(K^{\pm})}^{(n)} \ar@{.>}[ur]^{\exists ! \ \Delta} & }
\end{displaymath}
Following the same procedure we construct an algebra homomorphism
$\varepsilon: P_{B(K^{\pm})}^{(n)}\rightarrow P_{B(K^{\pm})}^{(n)}
\otimes P_{B(K^{\pm})}^{(n)}$ and an algebra antihomomorphism $S:
P_{B(K^{\pm})}^{(n)}\rightarrow P_{B(K^{\pm})}^{(n)} \otimes
P_{B(K^{\pm})}^{(n)}$ which are completely determined by their
values on the generators of $P_{B(K^{\pm})}^{(n)}$ (i.e.: the
basis elements of $k\langle b_{i}^{+}, b_{j}^{-}, K^{\pm}
\rangle)$.  Note that in the case of the antipode we start by
defining a linear map $S$ from $k\langle b_{i}^{+}, b_{j}^{-},
K^{\pm} \rangle$ to the opposite algebra $(P_{B(K^{\pm})}^{(n)}
\otimes P_{B(K^{\pm})}^{(n)})^{op}$, with values determined by
equation \eqref{eq:HopfPBK} and following the above described
procedure we end up with an algebra anti-homomorphism: $S:
P_{B(K^{\pm})}^{(n)}\rightarrow P_{B(K^{\pm})}^{(n)} \otimes
P_{B(K^{\pm})}^{(n)}$.  \\
Now it is sufficient to verify the rest of the Hopf algebra axioms
(i.e.: coassociativity of $\Delta$, counity property for
$\varepsilon$, and the compatibility condition which ensures us
that $S$ is an antipode) on the generators of
$P_{B(K^{\pm})}^{(n)}$. This can be done with straightforward
computations (see \cite{DaKa}).
\end{proof}
The above constructed algebra $P_{B(K^{\pm})}^{(n)}$, is an
ordinary Hopf algebra in the sense that the comultiplication is
extended to the whole of $P_{B(K^{\pm})}^{(n)}$ as an algebra
homomorphism :
$$
\Delta : P_{B(K^{\pm})}^{(n)}\rightarrow P_{B(K^{\pm})}^{(n)}
\otimes P_{B(K^{\pm})}^{(n)}
$$
where $P_{B(K^{\pm})}^{(n)} \otimes P_{B(K^{\pm})}^{(n)}$ is
considered as the tensor product algebra with the usual product:
$$
(a \otimes b)(c \otimes d) = ac \otimes bd
$$
for any $a,b,c,d \in P_{B(K^{\pm})}^{(n)}$ and the antipode
extends as usual as an algebra anti-homomorphism.

\section{Discussion}
It is interesting to see the relation between the above
constructed Hopf algebras $P_{B(g)}^{(n)}$ and
$P_{B(K^{\pm})}^{(n)}$.

 From the point of view of the structure, an obvious question
arises: While $P_{B(g)}^{(n)}$ is a quasitriangular Hopf algebra
through the $R$-matrix: $R_{g}$ given in eq.
\eqref{eq:nontrivRmatrcz2}, there is yet no suitable $R$-matrix
for the Hopf algebra $P_{B(K^{\pm})}^{(n)}$. Thus the question of
the quasitriangular structure of $P_{B(K^{\pm})}^{(n)}$ is open.

 An other interesting point, concerns the representations of
$P_{B(K^{\pm})}^{(n)}$ versus the representations of
$P_{B(g)}^{(n)}$. The difference in the comultiplication between
the above mentioned Hopf algebras, leads us to the question of
whether the tensor product of representations of $P_{B(g)}^{(n)}$
behave differently from the tensor product of representations of
$P_{B(K^{\pm})}^{(n)}$.

Finally another open problem which arises from the above mentioned
approach, is whether the above construction of
$P_{B(K^{\pm})}^{(n)}$ can be extended for the universal
enveloping algebra of an arbitrary Lie superalgebra, using power
series of suitably chosen
Casimirs.  \\ \\
 \textbf{Acknowledgements:} This paper is part of a
 project supported by ``Pythagoras II", contract number 80897.

\end{document}